# Dimensionality control and rotational symmetry breaking superconductivity in square-planar layered nickelates

Lin Er Chow,[1] Km Rubi,[2] King Yau Yip,[3] Mathieu Pierre,[4] Maxime Leroux,[4] Xinyou Liu,[3] Zhaoyang Luo,[1] Shengwei Zeng,[1] Changjian Li,[5] Michel Goiran,[4] Neil Harrison,[2] Walter Escoffier,[4] Swee Kuan Goh,[3] A. Ariando[1,✉]

[1]Department of Physics, Faculty of Science, National University of Singapore, Singapore 117551, Singapore

[2]National High Magnetic Field Laboratory, Los Alamos National Laboratory, Los Alamos, New Mexico 87545, USA

[3]Department of Physics, The Chinese University of Hong Kong, Shatin N.T., Hong Kong SAR, China

[4]LNCMI, Université de Toulouse, CNRS, INSA, UPS, EMFL, 31400 Toulouse, France

[5]Department of Materials Science and Engineering, Southern University of Science and Technology, Shenzhen, 518055, Guangdong, China

✉To whom correspondence should be addressed: ariando@nus.edu.sg

**The interplay between dimensionality and various phases of matter is a central inquiry in condensed matter physics.[1] New phases are often discovered through spontaneously broken symmetry.[2] Understanding the dimensionality of superconductivity in the high-temperature cuprate analogue – layered nickelates[3–8] and revealing a new symmetry-breaking state are the keys to deciphering the underlying pairing mechanism. Here, we demonstrate the highly-tunable dimensionality and a broken rotational symmetry state in the superconductivity of square-planar layered nickelates. The superconducting state, probed by superconducting critical current and magnetoresistance within superconducting transition under direction-dependent in-plane magnetic fields, exhibits a $C_2$ rotational symmetry which breaks the $C_4$ rotational symmetry of the square-planar lattice. Furthermore, by performing detailed examination of the angular dependent upper critical fields at temperatures down to 0.5 K with high-magnetic pulsed-fields, we observe a crossover from two-dimensional to three-dimensional superconducting states which can be manipulated by the ionic size fluctuations in the rare-earth spacer layer. Such a large degree of controllability is desired for tailoring strongly two/three-dimensional superconductors and navigating various pairing landscapes for a better understanding of the correlation between reduced dimensionality and unconventional pairing. These results illuminate new directions to unravel the high-temperature superconducting pairing mechanism.**

# Main text

The discovery of high-temperature (high-$T_c$) superconductivity in the layered cuprates[9] with quasi-two-dimensional electronic structure and superconductivity prompted the question of the correlation between reduced dimensionality and high-temperature pairing mechanism.[10] Since then, the enthusiasm to unlock the ingredients of high-$T_c$ pairing motivated the discoveries of superconductivity in layered materials such as ruthenates,[11] cobaltates,[12] and iron pnictides.[13] However, none shared the analogue two-dimensional high-$T_c$ cuprate-like pairing mechanism. Standing beside copper in the periodic table, nickel of $Ni^{1+}$ oxidation state in the infinite-layer nickelates RNiO$_2$ (R = rare-earth) consists of an analogous square-planar NiO$_2$ layered structure and $3d^9$ electronic structure with lifted orbital degeneracy which resembles the $Cu^{2+}$ state in the high-$T_c$ cuprates. As the ideal cuprate analogue, realising superconductivity in the layered nickelates[5–8,14] nearly two decades after its theoretical prediction[3,4] marked an important step in untangling the high-$T_c$ makeup of cuprates.

The reignited excitement in the field propelled further comparison between the newfound layered nickelates and the cuprates.[15–23] One of the most important questions is: whether the superconductivity in nickelates is two-dimensional like the cuprates,[24,25] or three-dimensional like the high-$T_c$ multiband iron pnictides?[26] Answering the question of the dimensionality of superconductivity is the next most critical step in building theoretical models (e.g., interlayer hopping, R-Ni hybridisation, Ni $d_{z^2}$ flat-band)[27–31] to decipher the pairing mechanism and related competing orderings in the strongly-correlated phase diagram.[32] The richness of cuprates' phase diagram begs for the discovery of new phases or ordering in the nickelates, which is typically achieved through the identification of spontaneous symmetry-breaking states. For example, a broken translational symmetry state (charge ordering) was reported in the undoped and underdoped regimes of the hole-doped nickelates.[33–35] However, a symmetry-breaking state has yet to be observed inside the superconducting dome, except for the broken global gauge symmetry and rotational symmetry in the phase factor of the order parameter, as expected in an unconventional superconductor family.[21,28] Searching for a symmetry-breaking state inside the superconducting dome is important to the understanding of the interplay between superconductivity and various orderings, especially if quantum critical transition exists in nickelates.

In this article, we investigate the dimensionality of nickelates' superconductivity by performing a detailed analysis of the angular-dependent upper critical fields $H_{c2}(\theta)$ in various hole-doped nickelates at temperatures down to 0.06 $T_c$ using high-magnetic pulsed fields. Our results show that neither a purely two-dimensional nor a three-dimensional description is correct. Instead, the dimensionality of layered nickelates is highly tunable and can undergo a crossover from two-dimensional to three-dimensional superconducting states as temperature decreases to the 0 K limit. The dimensionality of superconductivity in the layered nickelates can be manipulated through the ionic size fluctuations in the rare-earth spacer layer with different hole dopants. Therefore, the layered nickelate family is an ideal platform for investigating the relationship between dimensionality and superconducting pairing. In addition, we probe the in-plane anisotropy of the superconductivity by measuring the variation in the superconducting critical current and magnetoresistance within the superconducting transition in a direction-dependent in-plane magnetic field. A $C_2$ rotational symmetry is observed, which breaks the $C_4$ rotational symmetry of the lattice structure. Such a rotational symmetry-breaking superconducting state may indicate nematicity in the superconducting phase of the strongly-correlated hole-doped nickelates family.

## 2D to 3D dimensionality crossover in the superconducting states

Dimensionality of superconductivity in various hole-doped nickelates was investigated by performing polar angular-dependent upper critical fields $H_{c2}(\theta)$ using high pulsed magnetic field up to 55 T with a measurement geometry as shown in **Fig. 1a**. Details of the sample fabrication and characterization techniques are described in the Methods section and presented in the **Extended Data Fig. 1 and Fig. 2**. The $H_{c2}(\theta)$ at the lowest temperatures well below the superconducting transition $T_c$ are shown in **Figs. 1b-c** and **Extended Data Fig. 3**, while the $H_{c2}(\theta)$ at temperatures near the $T_c$ are shown in **Figs. 1d-f**. A cusp-like peak can be observed near $\theta = 90°$ ($H_{c2}^{\parallel}$) for both Sr-/Ca-doped La-nickelates (**Figs. 1d & f**) and Sr-doped Nd-nickelates (**Fig. 1e**). Such an anisotropic $H_{c2}(\theta)$ is typical for quasi-two-dimensional superconductivity, which can be described by the 2D-Tinkham model:

$$\left[\frac{H_{c2}(\theta)\sin(\theta)}{H_{c2}(90°)}\right]^2 + \left|\frac{H_{c2}(\theta)\cos(\theta)}{H_{c2}(0°)}\right| = 1.$$

All the $H_{c2}(\theta)$ near $T_c$ show a good agreement with the 2D-Tinkham model, as shown by the solid fit curves. The anisotropy ratio $\gamma = H_{c2}^{\parallel}/H_{c2}^{\perp}$ is typically large for a two-dimensional system; for example, $Bi_2Sr_2CaCu_2O_x$ has a very large $\gamma \sim 150$.[24] On the other hand, isotropic 3D-like superconductor such as $(Ba,K)Fe_2As_2$ has a small $\gamma \approx 2$ near $T_c$ and $\gamma \approx 1$ at the lowest temperatures.[26] The behaviour of nickelates, described by the strongly temperature dependent $\gamma$ in **Extended Data Fig. 4**, is dissimilar to both the two-dimensional cuprates and isotropic iron pnictides.

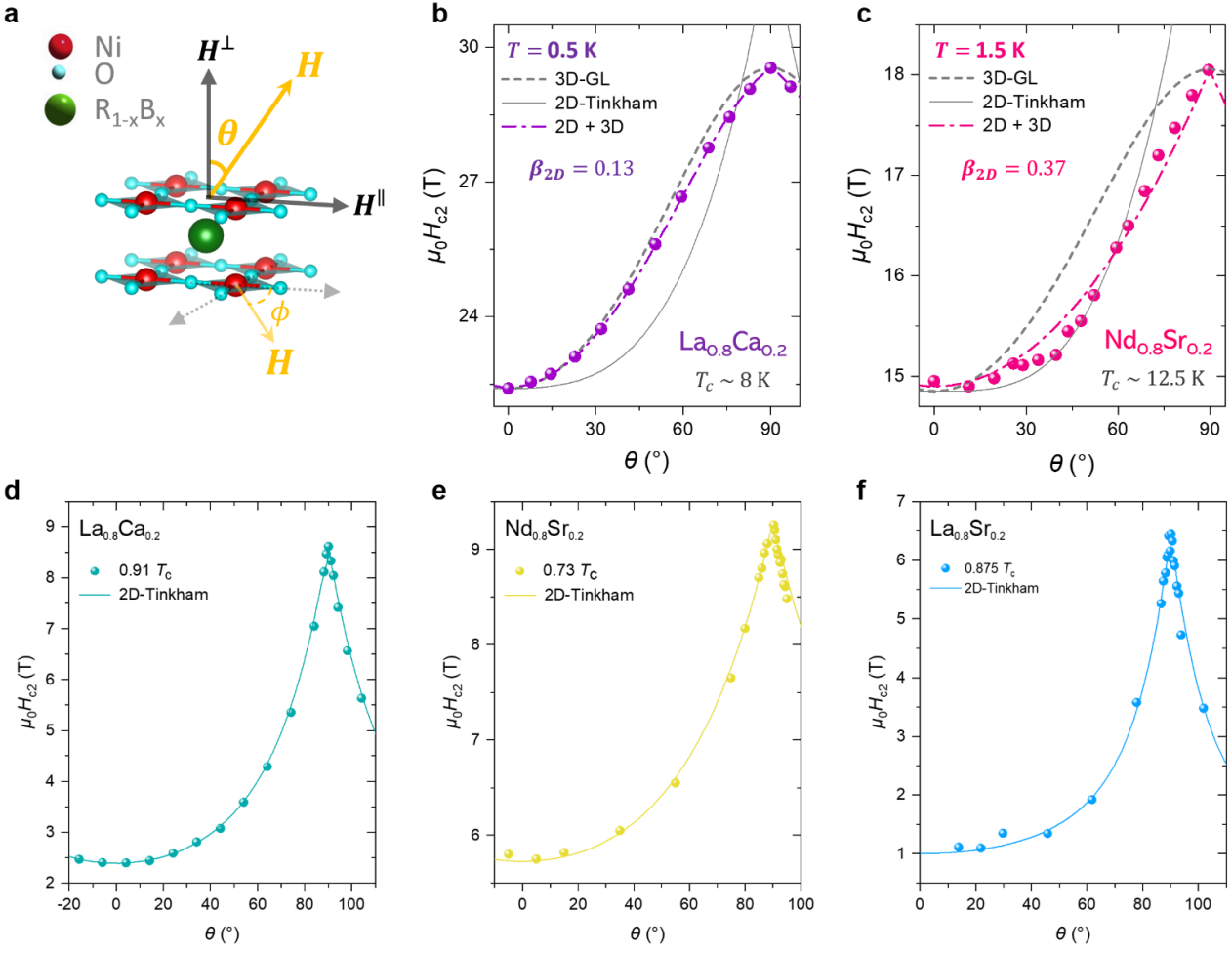

**Figure 1: Polar angular-dependent superconducting upper critical fields in layered nickelates.** The definition of field polar angle $\theta$ is schematically shown in (**a**) where R = La/Nd, B = Sr/Ca. (**b-c**) $\mu_0 H_{c2}(\theta)$ measured at the lowest temperatures of ~0.06 $T_c$ (**b**) and ~ 0.12 $T_c$ (**c**) using high pulsed magnetic fields up to 50 T (raw data shown in **Extended Data Fig. 2 & 3**). (**d-f**) $\mu_0 H_{c2}(\theta)$ measured near $T_c$ was fitted well to the 2D-Tinkham description. On the other hand, at the lowest temperatures (**b-c**), $\mu_0 H_{c2}(\theta)$ follows neither a purely 2D-Tinkham nor 3D-GL-like behaviour; instead, we can fit the data to $\left[\frac{H_{c2}(\theta)\sin(\theta)}{H_{c2}(90°)}\right]^2 + \alpha_{3D}\left[\frac{H_{c2}(\theta)\cos(\theta)}{H_{c2}(0°)}\right]^2 + \beta_{2D}\left|\frac{H_{c2}(\theta)\cos(\theta)}{H_{c2}(0°)}\right| = 1$ with $\alpha_{3D} + \beta_{2D} = 1$. For 2D-Tinkham, $\beta_{2D} = 1, \alpha_{3D} = 0$.

To reveal the temperature evolution of the dimensionality, we compare the $H_{c2}(\theta)$ near $T_c$ with those measured at the lowest possible temperatures of ~0.06 $T_c$. In contrast to the near $T_c$ behaviour, the upper critical fields of the Ca-doped LaNiO$_2$ (**Fig. 1b**) evolve smoothly as a function of angle $\theta$, suggesting a tendency towards a 3D-character at low temperatures. The magnitude of $H_{c2}$ is large and exceeds the Pauli-limit in all directions.[19] As shown in **Figs. 1b-c**, the $H_{c2}(\theta)$ of all systems cannot be fitted to both the 2D-Tinkham model and the 3D anisotropic mass Ginzburg-Landau (GL) model:

$$\left[\frac{H_{c2}(\theta)\sin(\theta)}{H_{c2}(90°)}\right]^2 + \left[\frac{H_{c2}(\theta)\cos(\theta)}{H_{c2}(0°)}\right]^2 = 1.$$

Instead, we fit the $H_{c2}(\theta)$ data to a combined 2D + 3D model:[36,37]

$$\left[\frac{H_{c2}(\theta)\sin(\theta)}{H_{c2}(90°)}\right]^2 + \alpha_{3D}\left[\frac{H_{c2}(\theta)\cos(\theta)}{H_{c2}(0°)}\right]^2 + \beta_{2D}\left|\frac{H_{c2}(\theta)\cos(\theta)}{H_{c2}(0°)}\right| = 1,$$

which measures the relative dominance of the 2D-Tinkham behaviour ($\beta_{2D}$) vs 3D-GL ($\alpha_{3D}$). We set $\beta_{2D} + \alpha_{3D} = 1$. When $\beta_{2D} \to 1$, a 2D-Tinkham description is observed. When $\beta_{2D} \to 0$, a 3D-GL model is fitted. Note that this equation is sensitive to the cusp behaviour near $\theta = 90°$ instead of the general anisotropy $\gamma$ value. For example, the Sr-doped Nd-nickelate in **Fig. 1e** has a small $\gamma \approx 1.6$, but the sharp peak near $\theta = 90°$ leads to a completely two-dimensional description which the $H_{c2}(\theta)$ is fitted well to the 2D-Tinkham model ($\beta_{2D} = 1$).

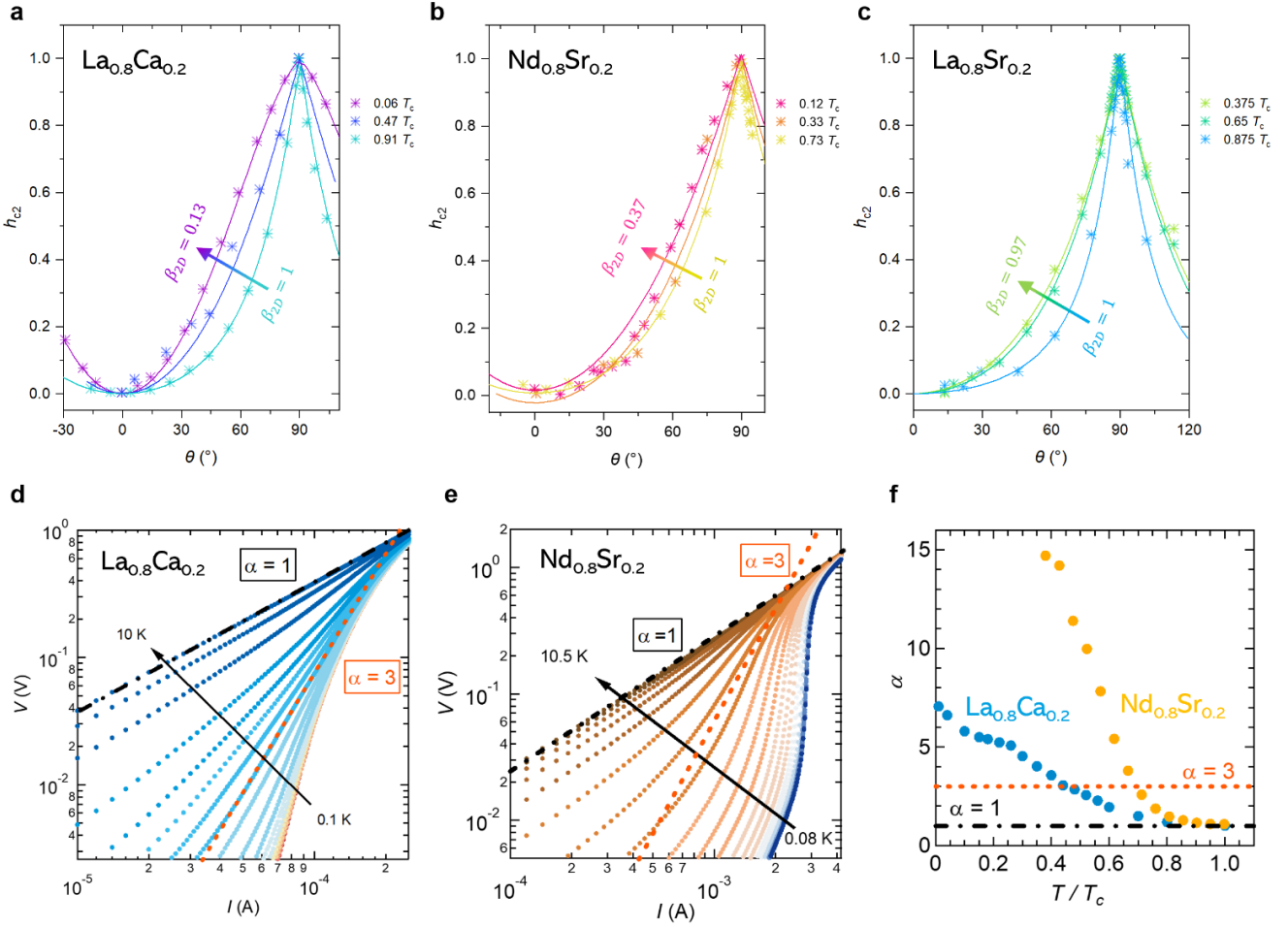

**Figure 2: Evidence for dimensionality tuning in infinite-layer nickelates.** A crossover from 2D-Tinkham-like to 3D-GL-like superconducting states can be observed in the Ca-doped LaNiO$_2$ (**a**) and Sr-doped NdNiO$_2$ (**b**). The parameter $h_{c2}(\theta) = \frac{H_{c2}(\theta) - H_{c2}^{\perp}}{H_{c2}^{\parallel} - H_{c2}^{\perp}}$, $0 \leq h_{c2}(\theta) \leq 1$ is used to compare the polar angular dependence of upper critical fields as temperatures lowered towards $T/T_c \to 0$. (**d-f**) Evidence for dimensionality control visualised from the BKT transition interpreted from the $I - V$ characteristics at various temperatures down to 0.08 K. When $V \propto I^3$ or $\alpha = 3$, $T_{BKT,(La,Ca)} \sim 0.5\, T_c$, $T_{BKT,(Nd,Sr)} \sim 0.7\, T_c$, implying that the Sr-doped NdNiO$_2$ is more two-dimensional-like than the Ca-doped LaNiO$_2$, consistent with the $H_{c2}(\theta)$ fitting results (**a-c**).

**Figure 2** summarises the $T/T_c$ evolution of the dimensionality in the layered nickelates family: Ca-doped La-nickelates, Sr-doped La-nickelates, Sr-doped Nd-nickelates. **Figures 2a-c** depict the $H_{c2}(\theta)$ at different $T/T_c$ spanning from near $T_c$ regime to low-temperature regime. To aid the visual comparison, we normalise the vertical axis to $h_{c2}(\theta) = \frac{H_{c2}(\theta) - H_{c2}^\perp}{H_{c2}^\parallel - H_{c2}^\perp}$, where $0 \leq h_{c2}(\theta) \leq 1$. One can see the striking differences in the dimensionality evolution in various rare-earth (La/Nd) and hole dopant (Sr/Ca) systems. For the Ca-doped La-nickelates (**Fig. 2a**), the $H_{c2}(\theta)$ evolves from a cusp-like peak near $\theta = 90°$ at temperatures near $T_c$ to a smooth peak at the lowest temperature of 0.06 $T_c$, demonstrating the '2D-ness' $\beta_{2D} = 1 \rightarrow \beta_{2D} = 0.13$ as $T \rightarrow 0$. Additional $H_{c2}(\theta)$ data is presented in **Extended Data Fig. 5**. However, for larger dopant size $Sr^{2+}$ as compared to the rare-earth ions, the Sr-doped system has a far stronger two-dimensional character in the superconducting states at the lowest temperature; for the Sr-doped LaNiO$_2$ (**Fig. 2c**), $H_{c2}(\theta)$ at temperature down to 0.375 $T_c$ still can be fully captured by the 2D-Tinkham behaviour. A consistent trend in the '2D-ness' of various rare-earth and hole dopant-based nickelates can be observed in the Berezinskii–Kosterlitz–Thouless (BKT) description of 2D superconductivity from the current-voltage characteristic plotted in logarithmic scale in **Figs. 2d-e**. The BKT transition can be taken at $\alpha = 3$ in $V \propto I^\alpha$. As shown in **Fig. 2f**, the Sr-doped NdNiO$_2$ still presents a relatively significant change in $\alpha$ as compared to the Ca-doped LaNiO$_2$ despite having a smaller NiO$_2$ plane separation (**Fig. 3**), and a defined $T_{BKT}$ closer to the $T_c$, which suggests the strong influence of dopant size (large $Sr^{2+}$ as compared to the rare-earth ions) in controlling the dimensionality in layered nickelates.

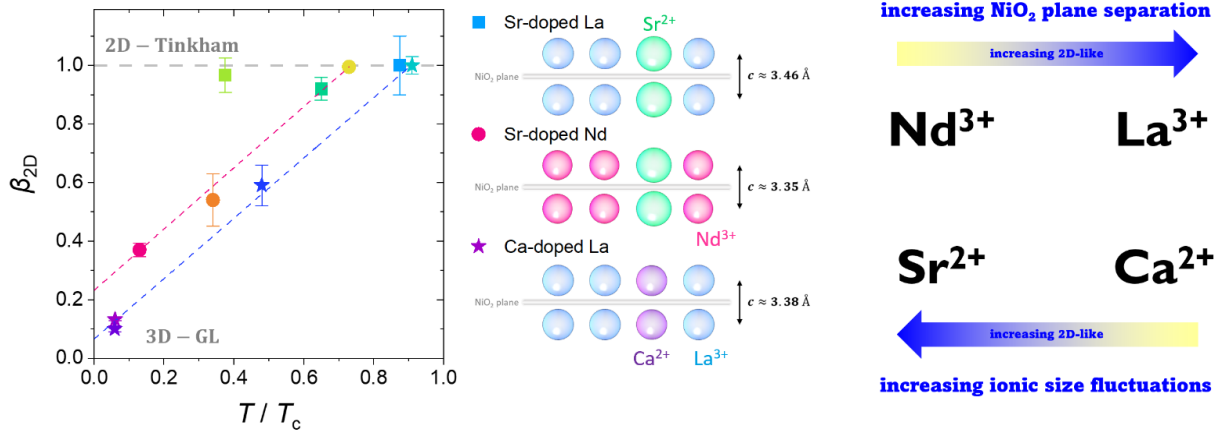

**Figure 3: Role of the rare-earth ions $Nd^{3+}$, $La^{3+}$ and hole dopants $Sr^{2+}$, $Ca^{2+}$ in controlling the dimensionality of infinite-layer nickelates.** The left figure plotted the fitted $\beta_{2D}$ value from $H_{c2}(\theta)$ of La$_{0.8}$Sr$_{0.2}$ (square), Nd$_{0.8}$Sr$_{0.2}$ (circle), La$_{0.8}$Ca$_{0.2}$ (star) at different $T/T_c$ down to the 0 K limit. The red and blue linear dashed lines are fitted to $\beta_{2D}\left(\frac{T}{T_c}\right) = \beta_0 + \beta'\left(\frac{T}{T_c}\right)$. At $T \to 0$ K limit, $\beta_0^{LaCa} = 0.066$, $\beta_0^{NdSr} = 0.23$. The small $\beta_0 \to 0$ suggests a dominant 3D-character of superconductivity. The slope $\beta' \sim 1$. While Ca-doped LaNiO$_2$ has a larger NiO$_2$ plane separation ($c \approx 3.38$ Å) than the Sr-doped NdNiO$_2$ ($c \approx 3.35$ Å), Ca-doped LaNiO$_2$ exhibits more 3D-like character at the lowest temperature suggesting the strong influence of ionic size fluctuations in the rare-earth spacer layer caused by hole dopants doped into the $A$-site layer. The middle panel depicts a schematic illustration of the ionic size fluctuations with $A$-site ions and $c$-axis lattice constant (= NiO$_2$ plane separation) drawn to scale: $r^{La^{3+}} \sim 117, r^{Nd^{3+}} \sim 112, r^{Ca^{2+}} \sim 114, r^{Sr^{2+}} \sim 132$ pm.

We summarise the 2D-to-3D dimensional crossover in the superconductivity of infinite-layer nickelates in **Fig. 3** and correspond the observation to the NiO$_2$ plane separation and the ionic size fluctuations in the rare-earth spacer layer caused by the size of the hole dopants $Sr^{2+}$, $Ca^{2+}$. The $\beta_{2D}$ value tracks the '2D-ness' of various nickelates. The first hypothesis should be the larger the NiO$_2$ plane separation, the more two-dimensional-like superconductivity in layered nickelates. This hypothesis is valid when comparing La-nickelate (larger $c \approx 3.46$ Å) and Nd-nickelate ($c \approx 3.35$ Å) with the same hole dopant Sr$^{2+}$, where the Sr-doped La-nickelate remains purely two-dimensional at ~ 0.38 $T_c$, while the Sr-doped Nd-nickelate has $\beta_{2D} \approx 0.6$, demonstrating a transition to partially three-dimensional in the superconducting coherence. However, when a smaller Ca$^{2+}$ is doped into the rare-earth spacer layer, even though the Ca-doped La-nickelate has a larger NiO$_2$ plane separation than the Sr-doped Nd-counterpart, superconductivity in the Ca-doped La-nickelate appears to be more three-dimensional in both the $H_{c2}(\theta)$ behaviour near $\theta = 90°$ and $I - V$ characteristic. Here we note that the conclusion of 2D-to-3D dimensional crossover superconductivity is consistent with the interpretation from the superfluid density near $T_c$ behaviour.[18]

We propose that while the first picture of increasing the NiO$_2$ plane separation can lead to a stronger two-dimensional character, the key to tuning the superconductivity in layered nickelates to be two-dimensional like the high-$T_c$ cuprates lies in modifying the spacer layer, where increasing the ionic size fluctuations in the rare-earth spacer layer can dramatically change the dimensionality of superconductivity in nickelates. With this observation, we can speculate that the Ba-doped LaNiO$_2$ is a purely two-dimensional superconductor with the highest hope of achieving cuprates-like pairing in infinite-layer nickelates. On the other hand, one can boost the 3D character and the corresponding responsible pairing mechanism by synthesizing Ca-doped NdNiO$_2$, which should carry the strongest 3D-like and potentially nodeless superconductivity in the layered nickelates family. The present work resolves the longstanding debates on whether nickelates' superconductivity is two-dimensional or three-dimensional and proposes that layered nickelates are the ideal platform to manipulate dimensionality and study the relationship between reduced dimensionality and superconducting pairing mechanism.

**Broken rotational symmetry in the superconducting states**

The anisotropy of the superconductivity in the square-planar NiO$_2$ planes was investigated by probing the change in the superconducting observables: (1) the superconducting critical current $I_c[B^\parallel(\phi)]$, which is related to the superconducting gap, (2) the fraction of the superconducting state relative to the normal state visualised in $\rho[B^\parallel(\phi)]$ within the superconducting transition, upon applying an in-plane magnetic field $B^\parallel$ along different directions $\phi$. **Figures 4a,d** show the in-plane azimuthal angular dependent $I_c[B^\parallel(\phi)]$ at temperatures from ~0.1 K to ~5 K in the Nd-nickelate Nd$_{0.8}$Sr$_{0,.2}$NiO$_2$ and La-nickelate La$_{0.8}$Ca$_{0,.2}$NiO$_2$. In both Nd- and La-nickelates, we see a two-fold modulation of the $I_c$, which breaks the four-fold $C_4$ rotational symmetry of the square-planar NiO$_2$ lattice structure. In the Nd-nickelates, in addition to the $C_2$ symmetric component, a $C_4$ symmetric component is present. To decouple the $C_2$ vs $C_4$ contributions, we fitted the $I_c[B^\parallel(\phi)]$ data to

$$A_{C2}\sin(2(\phi - \phi_{C2})) + A_{C4}\sin(4(\phi - \phi_{C4})) + A$$

as shown in the grey dashed lines. For La-nickelate, only the $C_2$ symmetric component is necessary to describe the in-plane variation in $I_c[B^\parallel(\phi)]$. For Nd-nickelate, we can fit the data to a significant $C_4$ component. The four-fold contributions exist at a broad temperature range but increase as temperature decreases from 5.5 to 0.15 K. Such temperature dependence of the $C_4$ rotational symmetry for the Nd-nickelates only can most naturally be explained as the orientation of the Nd$^{3+}$ $4f$ moments.

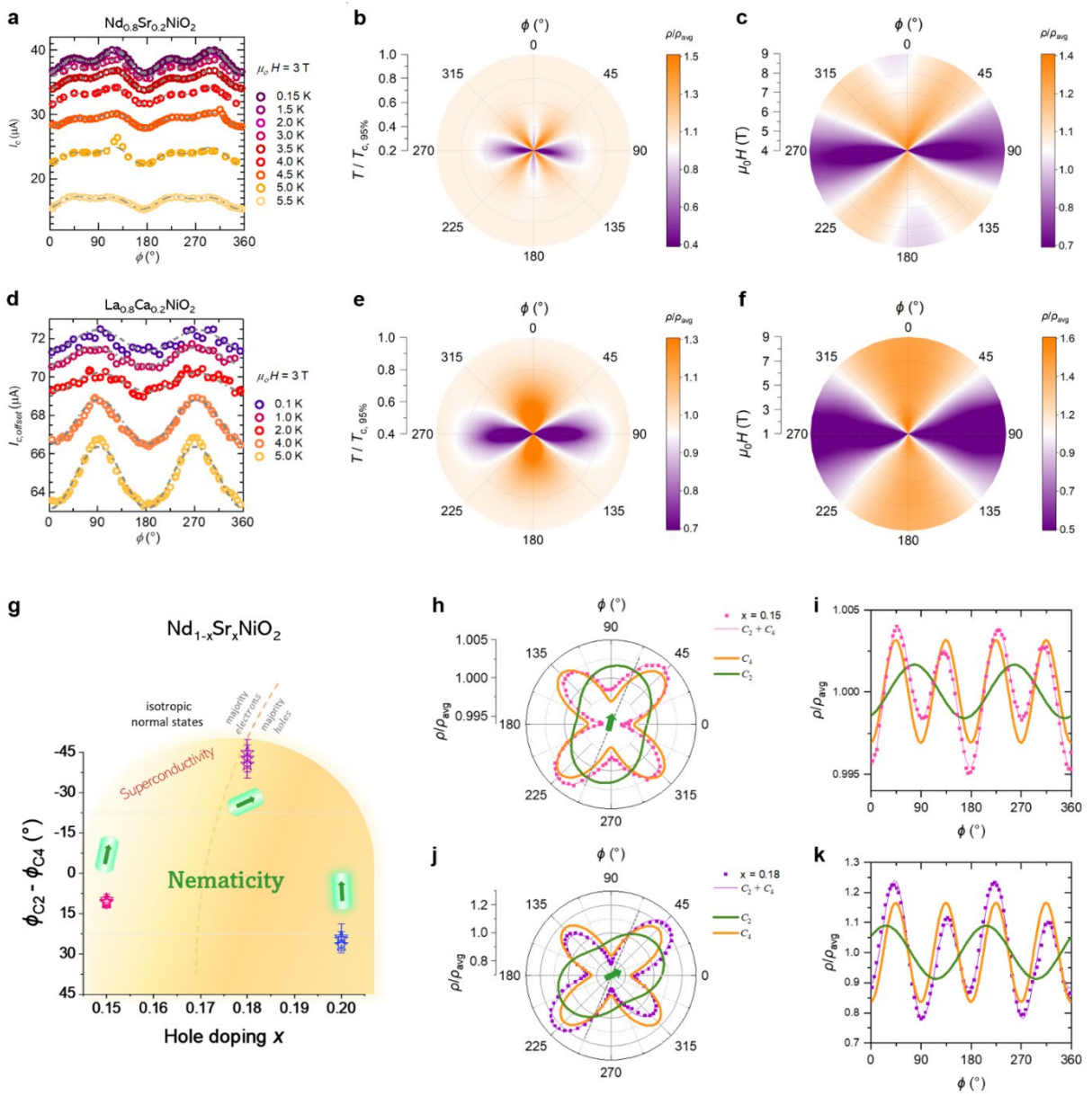

**Figure 4: Evidence for (in-plane) anisotropic superconducting observables in square-planar nickelates.** Representative Nd-nickelates $Nd_{0.8}Sr_{0.2}NiO_2$ (**a-c**) and La-nickelates $La_{0.8}Ca_{0.2}NiO_2$ (**d-f**). (**Top left: a,d**) Modulation of the superconducting critical current $I_c$ as a function of in-plane field $B^{\parallel}$ orientation measured using a 2D-vector magnet. Data for La-nickelate (**d**) are offset vertically for clarity. Sinusoidal functions of $C_2 + C_4$ symmetry $A_{C2}\sin(2(\phi - \phi_{C2})) + A_{C4}\sin(4(\phi - \phi_{C4}))$ show a good fit to the Nd- nickelate $I_c[B^{\parallel}(\phi)]$ (**a**) and magnetoresistance $\rho[B^{\parallel}(\phi)]$ within superconducting transition (**h-k**), while only a $C_2$ symmetry term is needed to describe the anisotropic $I_c[B^{\parallel}(\phi)]$ of La-nickelate (**d**). (**b-c, e-f**) Polar maps (in-plane azimuthal angle of $B^{\parallel}$) of the anisotropic

magnetoresistance at different temperatures $T/T_{c,95\%}$ (**b,e**) and different magnitudes of $B^{\parallel}$ (**c,f**). Data were measured at 9 T (**b,e**) and ~ 0.5 $T/T_{c,95\%}$ (**c,f**). For La-nickelate (**d-f**), only a $C_2$ symmetry was observed in the superconducting state. (**g-k**) Hole doping dependence of the $C_2$ symmetry phase shift $\phi_{C2}$ with respect to $\phi_{C4}$ in Nd$_{1-x}$Sr$_x$NiO$_2$. Data for representative $x = 0.15$ (**h-i**) was measured at 9 K with $B^{\parallel} = 9$ T, $x = 0.18$ (**j-k**) was measured at 2 K with $B^{\parallel} = 4$ T.

We further investigate the consistency of the observation of anisotropic superconducting critical current by measuring the in-plane angular dependent $\rho[B^{\parallel}(\phi)]$ within the superconducting transition. To determine if there is any anisotropic contribution from the normal state electrons, we measure the azimuthal angular-dependent magnetoresistance at a temperature above $T_c$ or sufficiently large magnetic field to be in the normal state as shown in **Extended Data Figs. 6a-f**. The normal state azimuthal angular dependent magnetoresistance shows virtually a circle in the polar plot with negligible anisotropy being observed. Therefore, we can exclude the trivial contribution of normal state electrons in the $\rho[B^{\parallel}(\phi)]$ within the superconducting transition and observe only the anisotropy in the superconducting states. **Extended Data Figs. 6d-k** show the in-plane azimuthal angular dependent $\rho[B^{\parallel}(\phi)]$ for Nd$_{1-x}$Sr$_x$NiO$_2$, La$_{1-x}$Ca$_x$NiO$_2$, and La$_{1-x}$Sr$_x$NiO$_2$ square-planar nickelates family. For La-nickelates, we can see a clear two-fold pattern in the superconducting states presented as an oval- or infinity-shape in the polar plot. For Nd-nickelate, an additional four-fold component is present at all hole doping $x$, which manifests as a cloverleaf pattern. The anisotropic $\rho[B^{\parallel}(\phi)]$ is summarised in the polar maps in **Figs. 4b-c** for $C_2 + C_4$ symmetry in Nd-nickelates, and **Figs. 4e-f** for $C_2$ symmetry in the La-nickelates.

Next, we investigate the relationship between the $C_2$ symmetric component and the $C_4$ symmetric component in the Nd-nickelates in terms of the phase shift $\phi_{C2} - \phi_{C4}$. The same forms of the sinusoidal functions are used in the fitting as previously discussed for the $I_c[B^\parallel(\phi)]$. As shown in the fitting results for representative hole doping levels $x = 0.15$ (**Figs. 4h-i**) and $x = 0.18$ (**Figs. 4j-k**), the phase shift for $C_2$ symmetric component is different for different doping levels, as visualised in the green arrow in **Figs. 4g-h & j**, which can be interpreted as the $C_2$ symmetric director rotated as hole doping crossing the majority electrons – majority holes transition (sign reversal in the Hall coefficients) in the superconducting dome of hole-doped Nd-nickelates.

Before we discuss the interpretation of the $C_2$ symmetric anisotropy in the family of square-planar nickelates, we first rule out a few trivial explanations for the observation. First, we performed the measurement in a different current direction $\phi_{\tilde{I}}$ and see if the $C_2$ symmetric component shifted with phase $\phi_{\tilde{I}}$ on the same sample. As shown in **Extended Data Fig. 7**, the representative top panel data is measured at $\phi_{\tilde{I}} = 0°$ while the bottom panel data is measured at $\phi_{\tilde{I}} = 45°$ at 9 T magnetic field in Nd$_{0.8}$Sr$_{0.2}$NiO$_2$. We can see the two-fold pattern in the bottom panel did not shift by $\phi_{\tilde{I}} = 45°$ with the current direction. Therefore, we can rule out Lorentz effect on the current as the cause of $C_2$ symmetric anisotropy. Secondly, we can exclude sample tilt as an explanation for the $C_2$ symmetric anisotropy. As shown in **Extended Data Fig. 8**, for the same sample mounted with the same in-plane orientation but intentionally tilted to a small angle of $\theta_{tilt} \sim 10°$, the measured azimuthal angular dependent $\rho[B^\parallel(\phi)]$ pattern is completely changed to two sharp dips at $\phi_{dip} \sim 162°$ and $\phi_{dip} \sim 342°$, which are at a different position from the original two-fold minima positions. In addition, as we discussed in the polar angular dependent $H_{c2}(\theta)$, at low temperatures, the $H_{c2}$ variation near the in-plane direction $H_{c2}^\parallel$ is smooth, and the superconducting state is mostly three-dimensional. Therefore, any small potential misalignment in the experiments would have a negligible effect on the in-plane anisotropy.

The observed $C_2$ rotational symmetry in the family of square-planar nickelates would suggest a rotational-symmetry-breaking superconducting state that is unexpected from the four-fold symmetry of the lattice structure, which could be interpreted as nematicity. Such nematic superconductivity is different from the nematicity observed in the normal state of electrons in the strongly-correlated system such as iron-based superconductor,[38] layered ruthenates $Sr_2RuO_4$ and $Sr_3Ru_2O_7$,[39,40] heavy-fermion $URu_2Si_2$,[41] and in the high-$T_c$ cuprates.[42,43] Considering nickelate is designed as a cuprate-analogue, it may not be surprising to find nematicity in the square-planar nickelates. We remain cautious to note that, if the hole dopants $Sr^{2+}/Ca^{2+}$ form a particular ordering upon substitution of the rare-earth ions, such superlattice structure (for example a chain) may potentially lead to some anisotropies in both the normal state and superconducting state. However, we did not observe any anisotropy in the normal state. Similar rotational symmetry breaking superconductivity was proposed in the flat-band twisted-bilayer-graphene,[44] few-layer $NbSe_2$,[45] and $Cu_xBi_2Se_3$.[46] Nematic superconductivity is associated with the two-dimensional $E$ representation with odd parity $p$-wave pairing which should possess topological nature that is different from the $B$ representation for $d$-wave in cuprates. Interestingly, unlike cuprates, the additional contribution from the Ni $d_{z^2}$ flat-band in nickelates favour spin-triplet $p$-wave pairing,[28] that was proposed theoretically to be significant at a sufficient hole doping level which can be observed in the superconducting dome given the self-doping effect. On the other hand, another calculation based on density functional theory (DFT) + many-body perturbation theory proposed that even undoped nickelates have prominent $d_{z^2}$ flat-band near the Fermi level which can give rise to van Hove singularities in the density of states and electronic nematicity.[47] The next question is how does the (gap/spin/electronic) nematicity interacts with the superconductivity especially near the quantum critical point (if exists) of the strongly-correlated hole-doped nickelates, which will be relevant for the pairing mechanism and order parameters of this newfound family of superconductors.

*Note*: During the preparation of this manuscript, we became aware of two reports on the in-plane magnetoresistance anisotropy,[48,49] and one report on the $T_c(\theta)$ in Sr-doped La-nickelate.[50]

## Methods

**Sample growth and preparation**

The infinite-layer nickelates $Nd_{1-x}Sr_xNiO_2$ and $La_{1-x}(Ca/Sr)_xNiO_2$ thin films were grown on $SrTiO_3$ (001) substrates using pulsed laser deposition and $CaH_2$ topotactic reduction process under conditions as previously reported.[18] The electron microscopy (STEM annular bright field and high-angle annular dark field) and electron energy loss spectroscopy characterisation are shown in **Extended Data Fig. 1**.

**Polar and Azimuthal angular-dependent magnetotransport measurement**

The wire connection for the electrical transport measurement was made by Al ultrasonic wire bonding. The transport measurements at temperatures down to 2 K or in magnetic field up to 14 T were performed using a Quantum Design Physical Property Measurement System with a rotator with an angle range from -10° to 370°. The onset of transitions $T_{c,95\%}$ is defined as the temperature at which the resistivity value reaches 95% of resistivity at 20 K, $\rho_{20K}$. Unless otherwise specified, $T_c$ and $H_{c2}$ in this article denotes the 50% resistivity transition criteria.

**High magnetic pulsed-field transport measurement**

The polar $\theta$ angular dependent of the upper critical fields (and magnetoresistance) were measured in pulsed-field $^3$He/$^4$He-cryostats. The La-nickelates were measured in magnetic field strength up to 65 T (~ 80 ms) and at temperatures down to ~ 0.5 K at the pulsed field facility of NHMFL in Los Alamos. The Nd-nickelates were measured in LNCMI-Toulouse with field strength up to 55 T (~ 300 ms) and temperature down to ~ 1.5 K.

**In-plane angular dependent superconducting critical current measurement**

The $I - V$ characteristics of the superconducting nickelates were measured in a $^3$He-$^4$He Bluefors LD250 dilution fridge with a base temperature of < 10 mK. To avoid sizeable Joule heating due to transport current, the critical current measurements were conducted at a temperature above 80 mK, where the dilution fridge has a cooling power of at least 160 µW. The dilution fridge is equipped with a two-dimensional vector magnet with a rotatable field of 3 T. To extract the azimuthal $\phi$ angular dependent of the superconducting critical current $I_c[B^{\parallel}(\phi)]$ under an in-plane magnetic field, we first measure the $I - V$ curves at a series of various angle $\phi$, temperature, and field magnitude (e.g. $\phi = 12°, T = 0.1 \text{ K}, B = 3 \text{ T}$), then plot $dV/dI$ vs $I$, and extract $I_c$ using 5% transition criteria.

## Data availability

The data that support the findings of this study are available from the corresponding author upon reasonable request.

## Competing interest declaration

The authors declare no competing financial interests.

## Authors contribution

## Acknowledgement

We acknowledge the helpful discussions with Hiroshi Yamase, Liang Si, and Karsten Held. This research is supported by the Ministry of Education (MOE), Singapore, under its Tier-2 Academic Research Fund (AcRF), Grant No. MOET2EP50121-0018, and Research Grants Council of the Hong Kong SAR, Grant No. A-CUHK 402/19. We acknowledge the support of LNCMI-CNRS, a member of the European Magnetic Field Laboratory (EMFL) under the proposal numbers TMS03-122 and TMS04-122. The research at NHMFL is supported by the National Science Foundation through NSF/DMR-1644779 and the State of Florida, the US Department of Energy "Science of 100 Tesla" BES program, and the Laboratory Directed Research and Development program of Los Alamos National Laboratory under project 20210320ER.

# Extended Data

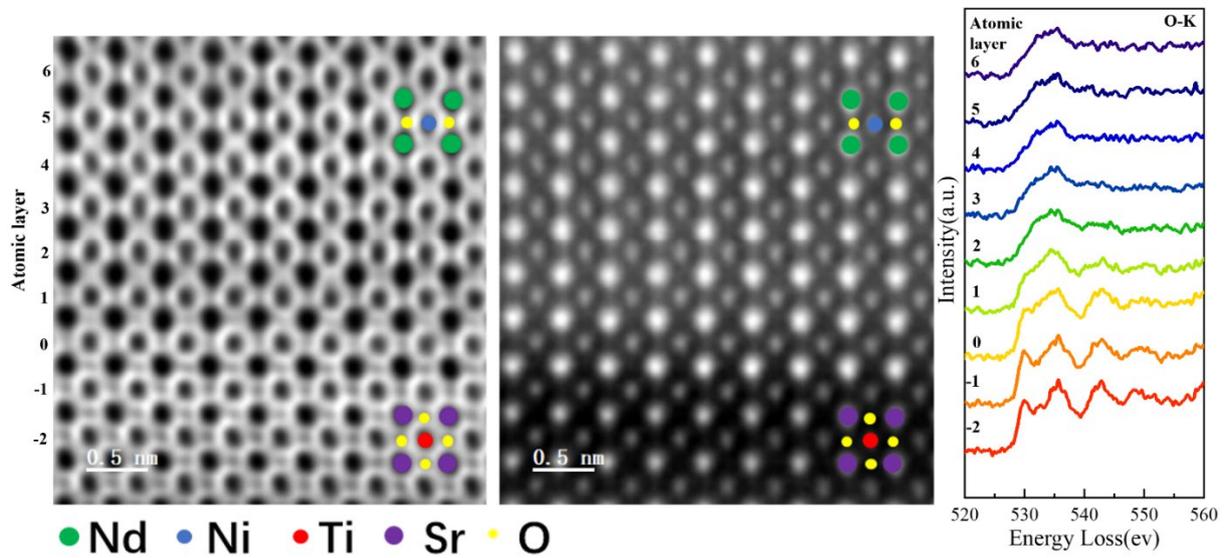

**Extended Data Figure 1: Electron microscopy and spectroscopy analysis of a representative infinite-layer nickelates $Nd_{0.8}Sr_{0.2}NiO_2$.** The interface layer between nickelates and $SrTiO_3$(001) substrate is denoted as layer '0'. From left to right: the annular-bright-field (ABF-STEM), high-angle annular dark-field (HAADF-STEM), and electron energy loss spectroscopy (EELS) at O $K$ edge.

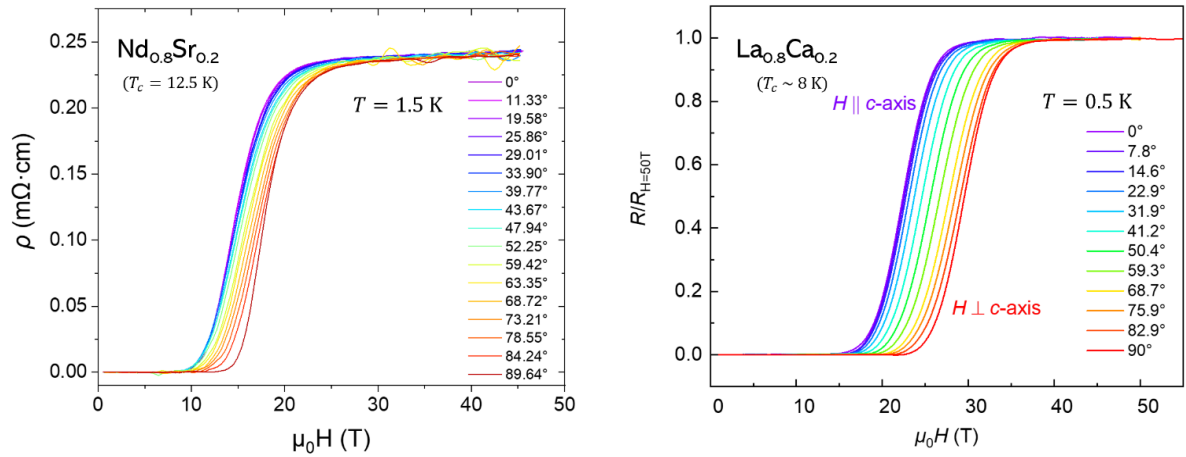

**Extended Data Figure 2: Magnetoresistance measured in high-magnetic pulsed-field setup at different polar angles $\theta$.** Representative low-temperature $R - H$ curves at different polar angles $\theta$ with respect to the out-of-plane ($\perp$ NiO$_2$ plane) direction for Nd$_{0.8}$Sr$_{0.2}$NiO$_2$ (**left**) and La$_{0.8}$Ca$_{0.2}$NiO$_2$ (**right**).

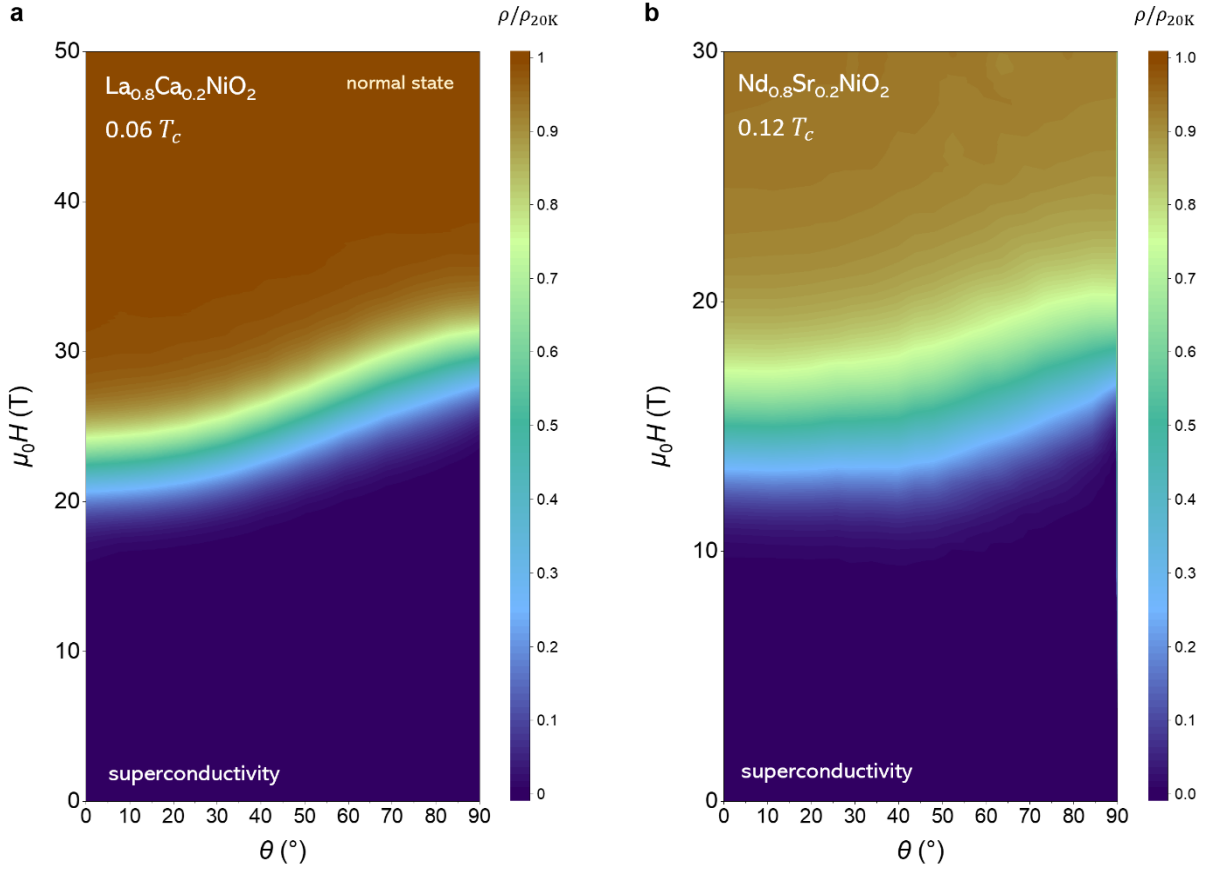

**Extended Data Figure 3: $H-\theta$ maps of the magnetoresistance measured in high-magnetic pulsed-field setup at different polar angles $\theta$.** Data is mapped with $R-H$ curves (shown in **Extended Data Fig. 2**) at different polar angles $\theta$ with respect to the out-of-plane ($\perp$ NiO$_2$ plane) direction for La$_{0.8}$Ca$_{0.2}$NiO$_2$ (**a**) and Nd$_{0.8}$Sr$_{0.2}$NiO$_2$ (**b**).

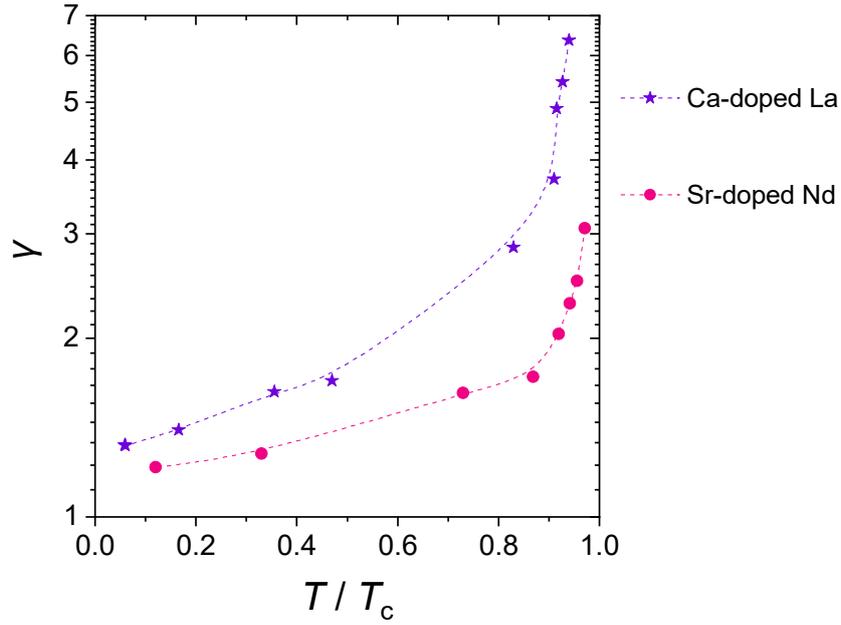

**Extended Data Figure 4: Two-dimensional superconductivity in infinite-layer nickelates near $T_c$.** The anisotropy ratio $\gamma = H_{c2}^{\parallel}/H_{c2}^{\perp}$ shows a diverging behaviour near $T_c$ while approaching < 2 at low temperatures. Note that for $Nd_{0.8}Sr_{0.2}NiO_2$ at 0.73 $T_c$ (see **Figure 1e**), $\gamma \sim 1.6$ is small but the $H_{c2}(\theta)$ shows a good fit to the 2D-Tinkham model. $Nd_{0.8}Sr_{0.2}NiO_2$ has a smaller $\gamma$ than $La_{0.8}Ca_{0.2}NiO_2$ at all temperature range; however, its $H_{c2}(\theta)$ shows a sharper cusp near $\theta = 90°$ (see **Figure 2a-b**) and suggests a more two-dimensional-like behaviour than the Ca-doped La-nickelates (see **Figure 3**) despite the smaller $\gamma$.

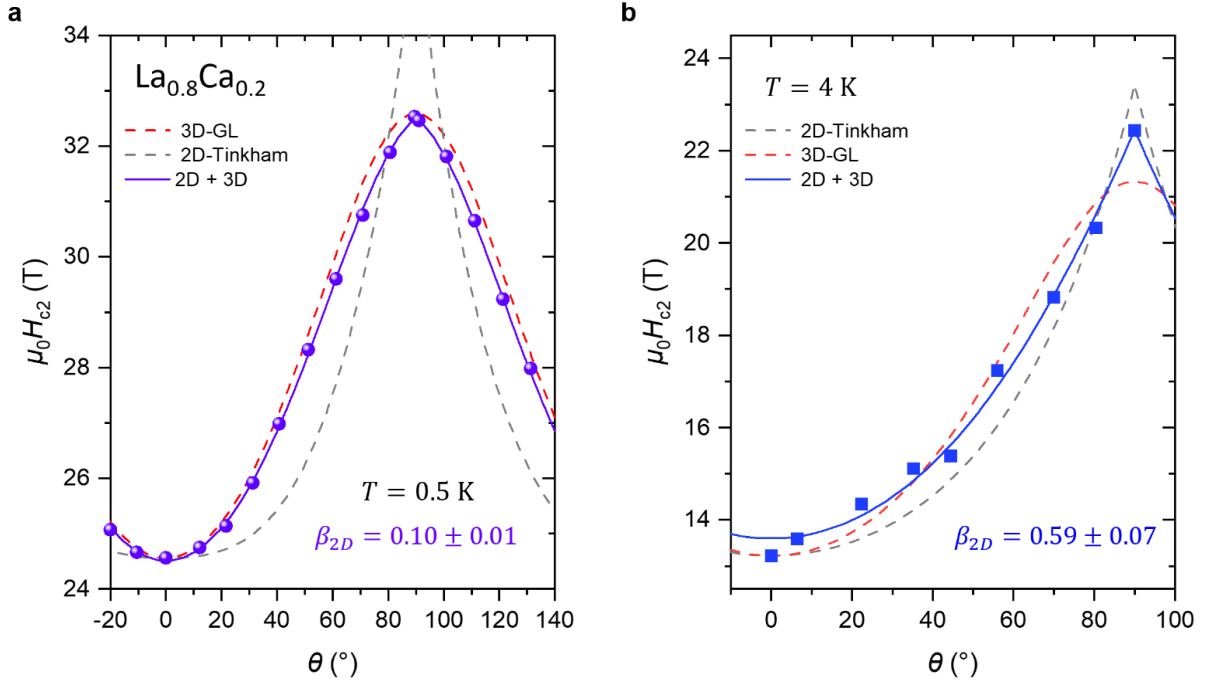

**Extended Data Figure 5: Additional set of data showing the crossover from 2D-Tinkham like to 3D-GL like superconducting states as $T \to 0$ K.** The representative data are measured in high-magnetic pulsed-field at 0.5 K (**a**) and 4 K (**b**), respectively, on La$_{0.8}$Ca$_{0.2}$NiO$_2$ ($T_c \approx 8$ K). The solid line 2D + 3D was fitted to the equations: $\left[\frac{H_{c2}(\theta)\sin(\theta)}{H_{c2}(90°)}\right]^2 + \alpha_{3D}\left[\frac{H_{c2}(\theta)\cos(\theta)}{H_{c2}(0°)}\right]^2 + \beta_{2D}\left|\frac{H_{c2}(\theta)\cos(\theta)}{H_{c2}(0°)}\right| = 1$. One can recover the 2D-Tinkham description by setting $\beta_{2D} = 1$, $\alpha_{3D} = 0$. Similarly, the 3D anisotropic mass Ginzburg–Landau model is recovered when $\alpha_{3D} = 1$, $\beta_{2D} = 0$.

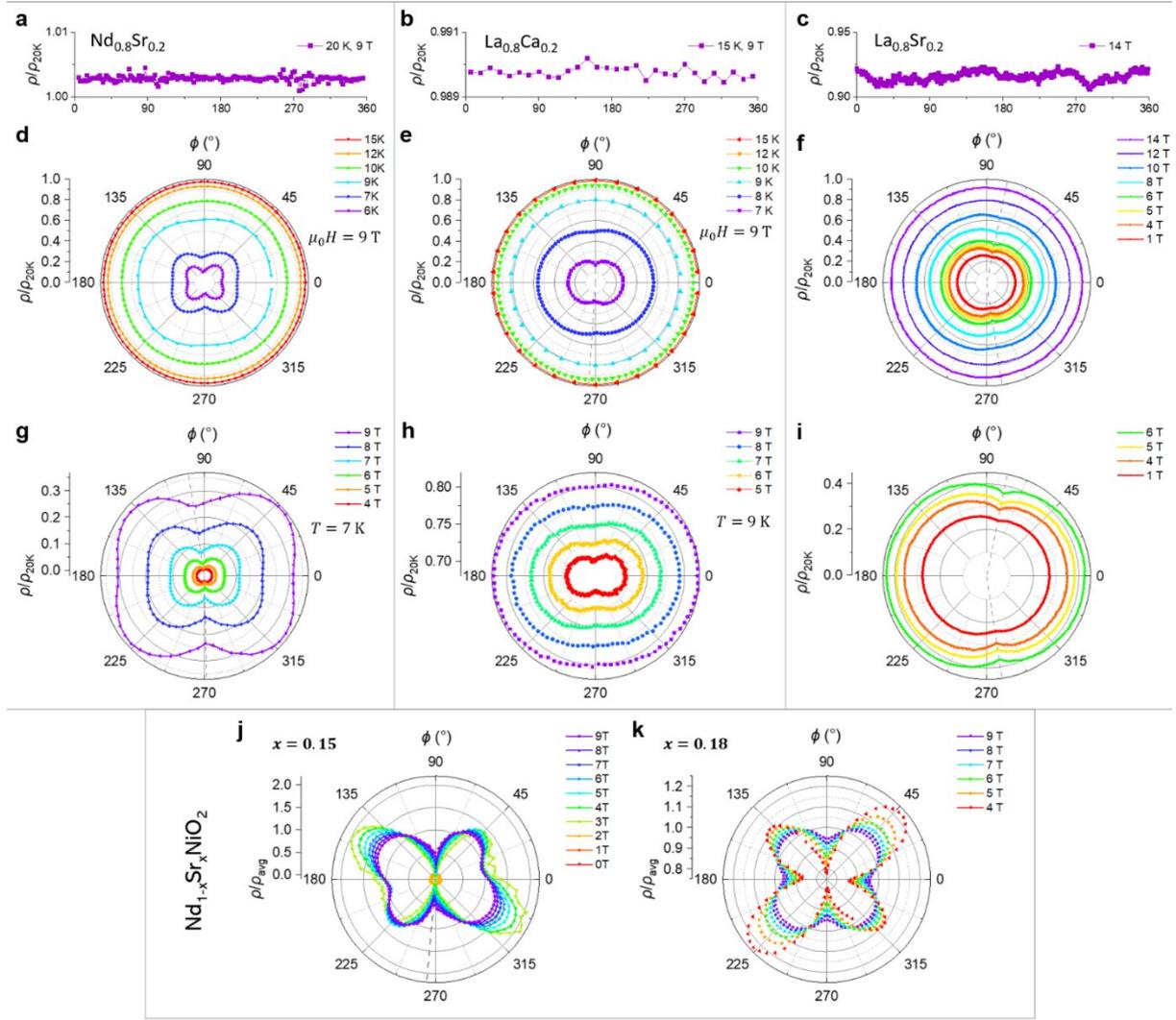

**Extended Data Figure 6: Anisotropic in-plane (azimuthal $\phi$) angular dependent magnetoresistance $\rho[B^{\parallel}(\phi)]$ within superconducting transition.** $\rho_{20K}$ is the zero-field resistivity at 20K. $\rho_{avg}$ is the average resistivity $\langle\rho[B^{\parallel}(\phi)]\rangle$ of various $0° \leq \phi \leq 360°$. (**Top a-c**) The normal state ($> T_{c,95\%}$ for **a-b**, $\sim > H_{c2}$ for **c**) shows no anisotropy in magnetoresistance as a function of in-plane field $B^{\parallel}(\phi)$. The virtually isotropic normal state $\rho[B^{\parallel}(\phi)]$ can be visualised as a circle in (**d-f**). Below $T_c$, a $C_2 + C_4$ symmetry can be observed in Nd$_{1-x}$Sr$_x$NiO$_2$ (**left d,g & j-k**), while only a $C_2$ symmetry can be observed in the La-nickelates: Ca-doped (**middle e,h**) and Sr-doped (**right f,i**). Slight asymmetric two-fold lobes (angle difference between minima are not 180°, maxima amplitude is unequal) can be observed in the La-nickelates (**i**).

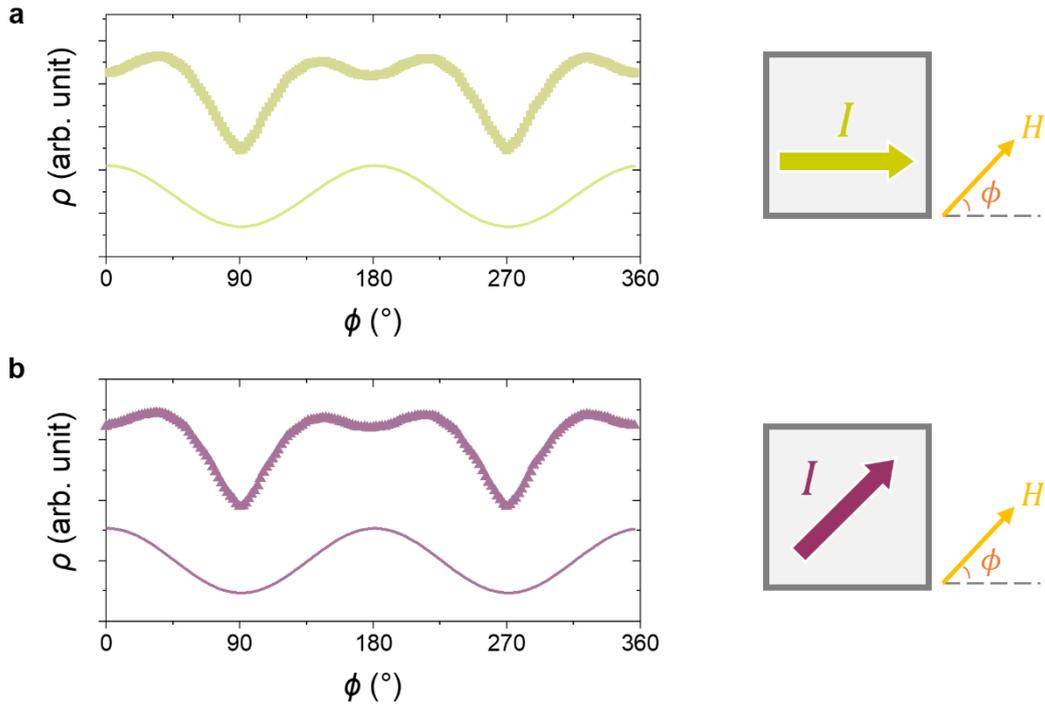

**Extended Data Figure 7: Anisotropic in-plane magnetoresistance when measured at two different current directions.** The in-plane magnetic field $B^{\parallel} = 9$ T. Representative data was measured at 10.5 K for $Nd_{0.8}Sr_{0.2}NiO_2$ ($T_{c,95\%} \sim 16.5$ K). (**a**) current $I$ direction points towards $\phi = 0°$. (**b**) current $I$ direction points towards $\phi = 45°$. We did not observe a corresponding 45° shift in the in-plane anisotropies. The observed $C_2$ symmetry was not originated from the Lorentz effect.

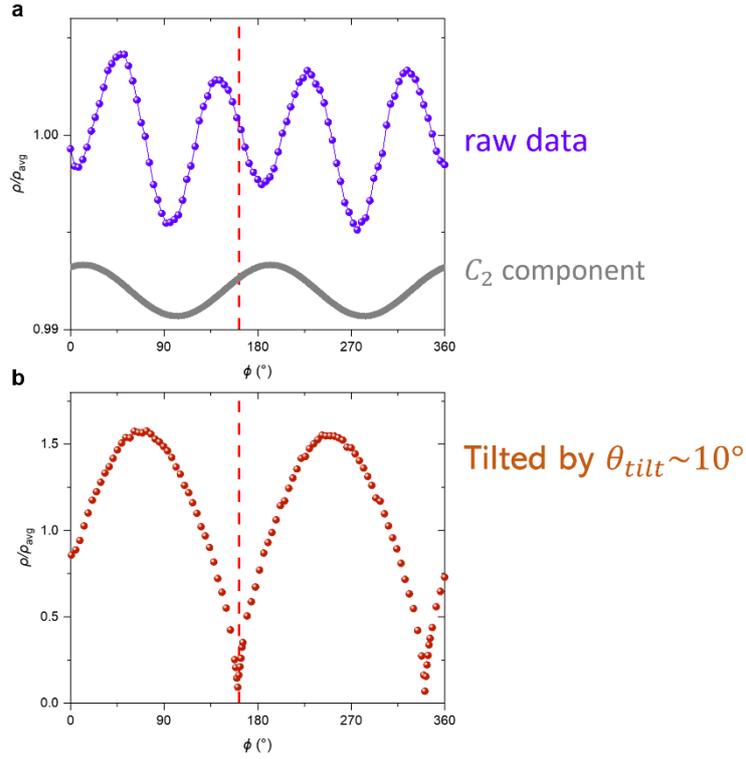

**Extended Data Figure 8: Excluding trivial origin of the anisotropic in-plane superconducting observables.** The $C_2$ symmetric component of $\rho[B^{\parallel}(\phi)]$ are not due to trivial misalignment of field angle (with $\theta_{tilt} \neq 0$). Note that the $H_{c2}(\theta)$ fitting results show that the superconductivity has a strong 3D character at low temperature for $Nd_{1-x}Sr_xNiO_2$ and $La_{1-x}Ca_xNiO_2$, and $H_{c2}(\theta)$ shows a smooth flattened peak near the in-plane angle $H \parallel NiO_2$ plane. Therefore, the tilt effect is negligible for our study. We can further prove that this is the case by intentionally introducing a small tilt angle of $\theta_{tilt} \sim 10°$ while maintaining the same in-plane orientation. Measurement on the angular dependence now shows a sharp dip (**b**) at a different angle $\phi_{dip} \sim 162°$ from the $C_2$ minima in (**a**). The grey $C_2$ component curve was shifted vertically for clarity. Representative data are measured at 9 T, 11 K (**a**) and 3 T, 8 K (**b**) in $Nd_{0.8}Sr_{0.2}NiO_2$ ($T_{c,95\%} \sim 14$ K).